%% file: root.tex
\documentclass[letterpaper, 10pt]{article}
\usepackage[margin=1in]{geometry}

\input{./Texts/00_Initialize}

\title{\LARGE
Pricing is All You Need to Improve Traffic Routing}

\author[1]{Yu Tang}
\author[1]{Kaan Ozbay}
\author[2]{Li Jin}
\affil[1]{\small C2SMARTER Center, Department of Civil \& Urban engineering, Tandon School of Engineering, New York University, 6 Metrotech Center, Brooklyn, New York, 11201, United States}
\affil[2]{\small UM Joint Institute and Department of Automation, Shanghai Jiao Tong University, 800 Dongchuan Lu, Shanghai, 200240, China}

\date{}

\begin{document}
\maketitle

\input{./Texts/10_Intro} 
\input{./Texts/20_Model} 
\input{./Texts/30_Analysis} 

\bibliographystyle{IEEEtran}
\bibliography{Bibliography}

\end{document}

%% file: Texts/00_Initialize.tex
\usepackage{amsmath}
\allowdisplaybreaks
\usepackage{xcolor}
\usepackage{footnote}
\usepackage{algorithm}
\usepackage{algpseudocode}
\usepackage{algorithmicx}
\usepackage{multirow} 
\usepackage{booktabs}
\usepackage{graphicx}
\usepackage{amssymb}
\usepackage{amsbsy}

\usepackage{authblk}

\usepackage{amsthm}

\usepackage{array}
\usepackage{longtable}
\usepackage{epstopdf}
\usepackage{pbox}
\usepackage{breqn}
\usepackage{mathrsfs}
\usepackage{multicol}
\usepackage{supertabular}
\usepackage{enumerate}
\usepackage[colorinlistoftodos]{todonotes}
\usepackage{url}
\usepackage[justification=centering]{caption}
\usepackage{tabu}
\usepackage{subcaption}
\usepackage{graphics} 

\usepackage{mathtools}

\usepackage{tikz}
\usepackage{tikz-network}

\usepackage{cases}
\usepackage{bm}
\usepackage{cite}

\usepackage{aligned-overset}

\newtheorem{thm}{Theorem}
\newtheorem*{thm*}{Theorem}

\theoremstyle{definition}

\newtheorem{dfn}{Definition}

\newcommand{\thistheoremname}{}
\newtheorem*{genericthm*}{\thistheoremname}
\newenvironment{namedthm*}[1]
  {\renewcommand{\thistheoremname}{#1}%
  \begin{genericthm*}}
  {\end{genericthm*}}

%% file: Texts/10_Intro.tex
\section{Introduction}

\subsection{Motivation}
Traffic routing control aims at reducing congestion via providing drivers with route guidance. Nevertheless, it has been reported that driver non-compliance with routing instructions could undermine the performance of this management strategy \cite{powell2000value}, especially social routing advice that deliberately detours part of vehicles to achieve benefits in terms of road networks \cite{van2019travelers}. Besides, our previous paper showed in a theoretical manner that driver non-compliance could destabilize routed traffic systems  \cite{tang2023does}. 

Fortunately, it is promising to secure traffic routing control via pricing strategies. This is because monetary costs also play an important role in route choices \cite{kerkman2012car}. Indeed, applying joint routing and pricing polices is not a new idea \cite{yang1999evaluating}. However, to the best of our knowledge, few studies have conducted an analytical analysis of these management approaches, particularly considering stochastic driver compliance influenced by congestion and tolls.   

\begin{figure}[htbp]
    \centering
    \begin{subfigure}{0.4\linewidth}
        \centering
        \begin{tikzpicture}
        \Vertex[x=0, style=black, size=0.1]{B}
        \Vertex[x=4, style=black, size=0.1]{N}
        \Edge[Direct,bend=30,lw=2pt](B)(N)
        \Edge[Direct,bend=330, lw=1pt](B)(N)
        \Edge[Direct,bend=315, lw=1pt](B)(N)
        \Edge[Direct,bend=300, lw=1pt](B)(N)
        \Text[x=2,y=0.9, fontsize=\footnotesize]{Corridor}
        \Text[x=2,y=-1.3, fontsize=\footnotesize]{Local streets}

        \Text[x=-0.5,y=-0.4, fontsize=\footnotesize]{Origin}
        \Text[x=4.8,y=-0.4, fontsize=\footnotesize]{Destination}
    \end{tikzpicture}
    \caption{A network consisting of one corridor and multiple local streets.}
    \end{subfigure}
    \begin{subfigure}{0.4\linewidth}
        \centering
        \begin{tikzpicture}
        \Vertex[x=-2.5,style={color=white}]{A}
        \Vertex[x=0, style=black, size=0.1]{B}
        \Vertex[x=4, style=black, size=0.1]{N}
        \Edge[Direct](A)(B)
        \Edge[Direct,bend=30](B)(N)
        \Edge[Direct,bend=330](B)(N)

        \Text[x=-1.25,y=0.2, fontsize=\footnotesize]{buffer $e_0$}
        \Text[x=2,y=0.9, fontsize=\footnotesize]{Corridor  $e_1$ with a fixed toll $p$}
        \Text[x=2,y=-0.9, fontsize=\footnotesize]{Local street  $e_2$}

        \Text[x=2,y=0.9, fontsize=\footnotesize]{}
        \Text[x=2,y=-1.3, fontsize=\footnotesize]{}
    \end{tikzpicture}
    \caption{An abstract of networks consisting of parallel links.}
    \end{subfigure}

    \caption{Modeling parallel-link networks.}
    \label{fig_twolink}
\end{figure}
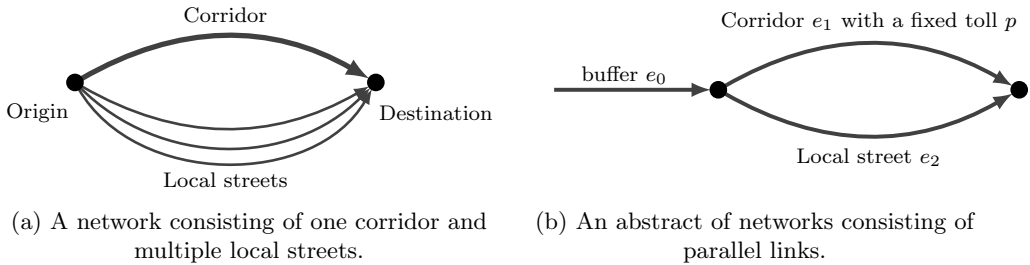

In this paper, we investigate the design of pricing policies that enhance driver adherence to route guidance, ensuring effective routing control. The major novelty lies in that we adopt a Markov chain to model drivers' compliance rates  conditioned on both traffic states and tolls. By formulating the managed traffic network as a nonlinear stochastic dynamical system, we can quantify in a more realistic way the impacts of driver route choices and thus determine appropriate tolls. Specially, we focus on a network comprised of two parallel links; see Figure~\ref{fig_twolink}. Though simple, the two-parallel-link network serves as a typical scenario for studying routing control; it turns out to be an appropriate abstraction of multiple parallel links: one stands for an arterial and the other denotes a set of local streets \cite{pi2017stochastic}. We assume that a reasonable routing policy is specified in advance, which means both the corridor $e_1$ and the local street $e_2$ are fully utilized if drivers completely obey the routing control. However, drivers could be reluctant to be detoured to link $e_2$. Thus a fixed toll $p$ is set on the corridor $e_1$ to give drivers incentives to choose the local street.

\subsection{Our contributions}
We try to address two main questions:
\begin{enumerate}
    \item[(i)] How to determine whether the network can be stabilized by routing and pricing strategies, subject to driver non-compliance?
    
    \item[(ii)] How to find the optimal pricing strategy that maximizes the network throughput, given a routing policy?
\end{enumerate}

The first question aims to assess the effectiveness of the given routing and pricing policies, where instability signals inadequate traffic management. To address this, we derive a stability condition (Theorem~\ref{thm_2}) using the Foster-Lyapunov criterion \cite{meyn2012markov} and an instability condition (Theorem~\ref{thm_3}) based on the transience of Markov chains \cite{meyn2012markov}.

It should be pointed out that the exact throughput cannot be determined, even for a simple two-parallel-link network, due to the randomness of driver non-compliance. To address the second question, we suggest using the stability and instability conditions to establish lower and upper bounds on throughput. This allows us to select suitable tolls that maximize these bounds.

%% file: Texts/20_Model.tex
\section{Problem Statement}
In this section, we first present our model formulation. Then we introduce the formal definitions of stability and throughput, which are closely related to our research problems. 

\subsection{Model formulation}

The considered network comprises links $e_0$, $e_1$ and $e_2$, as shown in Figure~\ref{fig_twolink}(b). Each link $e\in\{e_0,e_1,e_2\}$ is characterized by a length $l_e$ and a state of traffic density $x_e(t)$ at time $t$. Particularly, link $e_0$ serves a buffer receiving the upstream demand $D(t)\in\mathcal{D}$. We assume that $D(t)$ is governed by a stationary stochastic process with $\mathbb{E}[D(t)]=\Bar{D}$. Following the convention \cite{daganzo1995cell}, the buffer is assumed to have infinite storage, indicating $x_{e_0}(t)\in [0, \infty)$. Link $e_0$ is also associated with a bounded and non-decreasing sending flow $f_{e_0}:[0, \infty)\to[0, Q_{e_0}]$, where $Q_{e_0}$ denotes the capacity. We define its \emph{critical density} as 
\begin{equation}
    x_{e_0}^c:=\inf\{x|f_{e_0}(x)=Q_{e_0}\}, \label{eq_critical}
\end{equation}
which represents the lowest traffic density at which the sending flow $f_{e_0}$ attains its capacity. As for link $e\in\{e_1, e_2\}$, it only has finite storage such that $x_e\in[0, x_e^{\max}]$ where $x_e^{\max}$ is interpreted as the \emph{jam density}. In addition to the bounded and non-decreasing sending flow $f_{e}:[0, x_e^{\max}]\to[0, Q_e]$, link $e\in\{e_1, e_2\}$ also has a bounded and non-increasing receiving flow $r_{e}:[0, x_e^{\max}]\to[0, Q_e]$. For notational convenience, we let 
$\mathcal{X}:=[0,\infty)\times[0, x_{e_1}^{\max}]\times[0, x_{e_2}^{\max}]$ and $x:=[x_{e_0}, x_{e_1}, x_{e_2}]^{\mathrm{T}}\in\mathcal{X}$.

For traffic management strategies, we first consider a routing policy, denoted by $\alpha:\mathcal{X}\to[0,1]$, specifying the proportion of vehicles assigned onto link $e_1$. The routing policy $\alpha(x)$ is assumed to be non-increasing with respect to $x_{e_1}$ and non-decreasing with respect to $x_{e_2}$. Besides, a fixed toll $p\geq0$ is set on link $e_1$ for each passing vehicle. Then, we model drivers' response to the traffic management. Particularly, we denote by $C_e(x,p)$ the compliance rate regarding the routing instruction to link $e\in\{e_1,e_2\}$, which generally depends on the traffic state $x$ and the fee $p$. We consider that $C(x,p):=[C_{e_1}(x,p), C_{e_2}(x,p)]^{\mathrm{T}}$ is a random vector conditioned on $x$ and $p$, with a distribution $\Gamma_{x,p}$ supported on $\mathcal{C}\subseteq[0,1]^2$. We also assume i) that $\mathbb{E}[C_{e_1}(x,p)]$ is non-increasing with respect to $x_{e_1}$ and $p$, and non-decreasing with respect to $x_{e_2}$, and ii) that $\mathbb{E}[C_{e_2}(x,p)]$ is non-decreasing with respect to $x_{e_1}$ and $p$, and non-increasing with respect to $x_{e_2}$.

The conservation law yields the following dynamics:
\begin{subequations}
    \begin{align}
        x_{e_0}(t+1) =& x_{e_0}(t) + \frac{\delta_t}{l_{e_0}}\Big(D(t)-q_{e_1}(x,p,C)-q_{e_2}(x,p,C)\Big), \label{eq_sys_1} \\
        x_{e}(t+1) =& x_{e}(t) + \frac{\delta_t}{l_{e}}\Big(q_{e}(x,p,C)-f_{e}(x_e)\Big),~ e\in\{e_1, e_2\}, \label{eq_sys_2}
    \end{align}
\end{subequations}
where $\delta_t$ is the time step size, and the flow from link $e_0$ to link $e\in\{e_1, e_2\}$, denoted by $q_{e}(x, p, C)$, is given below. 
\begin{subequations}
    \begin{align}
            q_{e_1}(x, p, C) =& \min\Big\{\Big(\alpha(x)C_{e_1}(x,p) + (1-\alpha(x))(1-C_{e_2}(x,p))\Big)f_{e_0}(x_{e_0}), r_{e_1}(x_{e_1})\Big\}, \label{eq_q_1} \\
            q_{e_2}(x, p, C) =& \min\Big\{\Big(\alpha(x)(1-C_{e_1}(x,p)) + (1-\alpha(x))C_{e_2}(x,p)\Big)f_{e_0}(x_{e_0}), r_{e_2}(x_{e_2})\Big\}. \label{eq_q_2}
    \end{align}
\end{subequations}
Clearly, we obtain a nonlinear stochastic system \eqref{eq_sys_1}-\eqref{eq_sys_2} that is a Markov chain. 

Now we briefly discuss how the toll $p$ influences the inter-link flows $q_{e_1}(x, p, C)$ and $q_{e_2}(x, p, C)$. As mentioned in previous section, this paper considers that drivers may resist being redirected to local streets. When the toll $p$ is low, the compliance rate $C_{e_1}(x,p)$ is high but $C_{e_2}(x,p)$ could be low, consequently compromising the routing policy $\alpha$. However, extremely high tolls may render low $C_{e_1}(x,p)$ and high $C_{e_2}(x,p)$, which also nullifies traffic routing.  

\subsection{Stability and throughput}
The following gives the definition of stability considered in this paper. 
\begin{dfn}[Stability \& Instability] \label{dfn_sta}
A stochastic process $\{Y(t):t\geq0\}$ with a state space $\mathcal{Y}$ is \emph{stable} if there exists a scalar $Z<\infty$ such that for any initial condition $y\in\mathcal{Y}$
\begin{equation}\label{eq_bounded}
  \limsup_{t\to\infty}\frac{1}{t}\sum_{\tau=0}^t\mathbb {E}[|Y(\tau)||Y(0)=y] \le Z,
\end{equation}
where $|Y(\tau)|$ denotes 1-norm of $Y(\tau)$. The network is \emph{unstable} if there does not exist $Z<\infty$ such that \eqref{eq_bounded} holds for any initial condition $y\in\mathcal{Y}$.
\end{dfn}

The stability above is widely used in studying traffic control \cite{barman2023throughput}. It indicates that the time-average traffic density is bounded in the long term. Obviously, in practice one is more concerned about traffic performance within finite time (e.g. peak hours). Although Definition~\ref{dfn_sta} simplifies the analysis of real-world traffic systems, our later numerical examples illustrate that methods based on this definition are sufficient to produce insightful results for evaluating and designing management strategies. Moreover, this establishes a foundation for future research on refined finite-time stability \cite{hong2010finite}.

The \emph{throughput} $\Bar{D}^{\alpha,p}$ of the network, given the routing policy $\alpha$ and the toll $p$, is defined as the maximal expected demand that the network can accept while maintaining stability: 
\begin{equation*}
    \Bar{D}^{\alpha,p}:=\sup\Bar{D}\quad \text{subject to the system \eqref{eq_sys_1}-\eqref{eq_sys_2} is stable.} 
\end{equation*}

Our research problems are i) how to verify whether the system \eqref{eq_sys_1}-\eqref{eq_sys_2} under the routing and pricing policies satisfies the condition \eqref{eq_bounded}, and ii) how to select appropriate $p$ to maximize the throughput. 

%% file: Texts/30_Analysis.tex
\section{Major Results}
In this section, we present both theoretical and numerical results. We begin by introducing theorems that establish stability and instability conditions, followed by their application in stability verification. Next, we provide examples illustrating the alignment of our theorems with numerical simulations. We also discuss the insights gained from our proposed methods. 

\subsection{Stability \& instability conditions}

We present two theorems below. Theorem~\ref{thm_2} states one stability condition, derived using the Foster-Lyapunov criterion \cite{meyn2012markov}, while Theorem~\ref{thm_3} provides one instability condition, based on the transience property of Markov chains \cite{meyn2012markov}.
\begin{thm}
\label{thm_2}
The system \eqref{eq_sys_1}-\eqref{eq_sys_2} is stable if there exists a vector $\theta:=[\theta_{e_1},\theta_{e_2}]^{\mathrm{T}}\in[0, 1]^2$ and a negative scalar $\gamma<0$ such that
\begin{align}
    &\bar{D}- \sum_{e\in\{e_1,e_2\}} (1-\theta_{e})\mathbb{E}[q_{e}(x, p, C)]- \sum_{e\in\{e_1,e_2\}}\theta_e f_e(x) < \gamma, ~\forall x \in\{x\in\mathcal{X}|x_{e_0}=x_{e_0}^c\}, 
    \label{eq_thm2_1}
\end{align}
where $x_{e_0}^c$ is given by \eqref{eq_critical} and $\mathbb{E}[q_{e}(x, p, C)] := \int_{\mathcal{C}} q_{e}(x, p, c)) \Gamma_{x,p}(\mathrm{d}c)$.
\end{thm}

\begin{thm}
\label{thm_3}
The system \eqref{eq_sys_1}-\eqref{eq_sys_2} is unstable if there exists a vector $\theta:=[\theta_{e_1},\theta_{e_2}]^{\mathrm{T}}\in[0,1]^2$ and a non-negative scalar $\gamma\geq0$ such that
\begin{align}
    &\bar{D}- \sum_{e\in\{e_1,e_2\}} (1-\theta_{e})\mathbb{E}[q_{e}(x, p, C)] - \sum_{e\in\{e_1,e_2\}}\theta_{e} f_{e}(x) \geq \gamma, ~\forall x \in\{x\in\mathcal{X}|x_{e_0}=x_{e_0}^c\}. \label{eq_thm3_1}
\end{align}
\end{thm}

Theorem~\ref{thm_2} (resp. Theorem~\ref{thm_3}) essentially says that the network is stable (resp. unstable) if the weighted expected net flow is negative (resp. non-negative) over the traffic state space $\{x\in\mathcal{X}|x_{e_0}=x_{e_0}^c\}$. One can implement Theorem~\ref{thm_2} by solving the following Semi-Infinite Programming (SIP \cite{stein2012solve}):
\begin{equation*}
    (P_1)~ \min_{\theta,\gamma}~\gamma~\text{subject to}~\eqref{eq_thm2_1}.
\end{equation*}
If the optimal $\gamma^*$ is negative, the stability is concluded. Similarly, the instability verification requires solving the SIP: 
\begin{equation*}
    (P_2)~ \max_{\theta,\gamma}~\gamma~\text{subject to}~\eqref{eq_thm3_1}.
\end{equation*}
If the optimal $\gamma^*$ is non-negative, we say that the system \eqref{eq_sys_1}-\eqref{eq_sys_2} is unstable.

\subsection{Numerical examples}

The following presents settings in our numerical examples. First, the demand $D(t)$ is assumed to be uniformly distributed on $[D^{\min}, D^{\max}]$. Then, the sending and receiving flows are specified by $f_{e}(x) = \min\{v_{e}x_{e}, Q_{e}\}$ for $e\in\{e_0, e_1, e_2\}$ and $r_{e}(x) = \min\{R_{e}-w_{e}x_{e},  Q_{e}\}$ for $e\in\{e_1, e_2\}$, respectively. 
We consider a fixed routing ratio based on the link capacities $Q_{e_1}$ and $Q_{e_2}$, namely  $\alpha:=Q_{e_1}/(Q_{e_1}+Q_{e_2})$. The compliance rate $C_e(x,p)$ is uniformly distributed on  $[\max\{\bar{C}_e(x, p)-\epsilon_e, 0\}, \min\{\bar{C}_e(x, p)+\epsilon_e, 1\}]$, where $\bar{C}_e(x, p)$ is given by
\begin{equation*}
    \bar{C}_e(x, p) := \frac{1}{1+e^{\beta_{e}^0+\beta_{e}^1 x_{e_1} + \beta_{e}^2 x_{e_2} + \beta_{e}^3 p}}.    
\end{equation*}
The parameters are summarized in Table~\ref{tab_para}. Note that negative $\beta_{e_1}^0$ and positive $\beta_{e_2}^0$ indicate that drivers naturally prefer the corridor.
\begin{table}[htbp]
    \centering
    \caption{Parameter settings.}
    \small
    \begin{tabular}{ cc|cc|cc|cc|cc }
    \hline
    $v_{e_0}$ & 80 (km/h)   & $v_{e_1}$ & 100 (km/h)  & $v_{e_2}$ & 50 (km/h)  & $\beta_{e_1}^0$ & $-4$ & $\beta_{e_2}^0$ & $1$ \\ 
    $Q_{e_0}$ & 8000 (veh/h) & $Q_{e_1}$ & 4000 (veh/h) & $Q_{e_2}$ & 2000 (veh/h)  & $\beta_{e_1}^1$ & 0.01 & $\beta_{e_2}^1$ & $-0.02$ \\ 
     $D^{\min}$  & 4000 (veh/h)     & $R_{e_1}$ &  4800 (veh/h)    & $R_{e_2}$ & 2400 (veh/h) &  $\beta_{e_1}^2$ & $-0.02$ & $\beta_{e_2}^2$ & $0.03$ \\ 
     $D^{\max}$      & [5000, 8000] (veh/h)  & $w_{e_1}$ & 20 (km/h)   & $w_{e_2}$ & 10 (km/h)  & $\beta_{e_1}^3$ & $0.3$ & $\beta_{e_2}^3$ & $-0.6$ \\ 
           & & & & & & $\epsilon_{e_1}$ & 0.1 & $\epsilon_{e_2}$ & 0.1  \\
    \hline
    \end{tabular}
    \label{tab_para}
\end{table}

\subsubsection{Impacts of tolls and demands}
\begin{figure}[htbp]
    \centering
    \begin{subfigure}{0.25\linewidth}
        \centering
        \includegraphics[width=\linewidth]{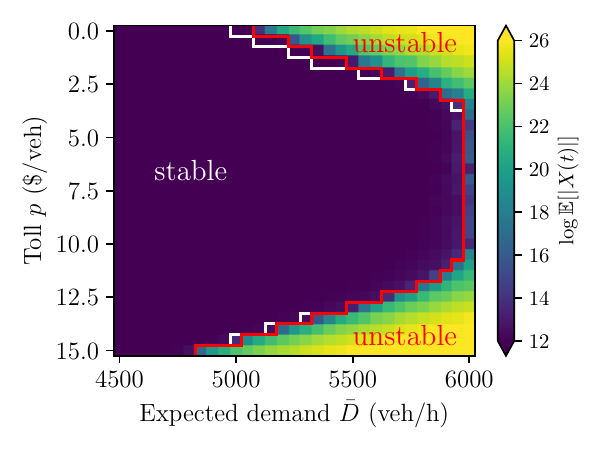}
    \caption{Stability regions.}
    \end{subfigure}
    \begin{subfigure}{0.25\linewidth}
        \centering
        \includegraphics[width=\linewidth]{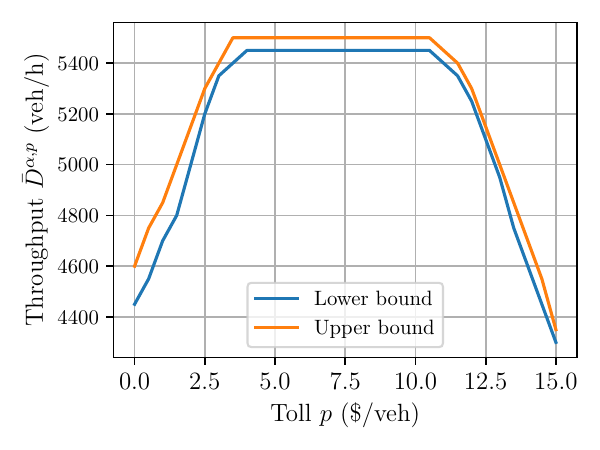}
    \caption{Throughput bounds.}
    \end{subfigure}

    \caption{Impact analysis of tolls and expected demands.}
    \label{fig_demand_pricing}
\end{figure}

Figure~\ref{fig_demand_pricing}(a) shows the time-average traffic densities after $10^4$ steps and reveals the stability and instability regions. The white boundary is obtained from Theorem~\ref{thm_2}, while the red boundary is derived from Theorem~\ref{thm_3}. Therefore, we can conclude that the region to the left of the white boundary is stable, whereas the areas in the upper right and lower right corners are unstable. These findings are consistent with the numerical results. Note that there is a gap between the white and red boundaries, within which stability or instability cannot be determined. This is because we do not have sufficient and necessary stability conditions. In fact the gap is not a concern, as it can be narrowed by using more advanced Lyapunov or test functions, though at the expense of increased computational cost. 

The key findings from Figure~\ref{fig_demand_pricing} are summarized as follows. First, setting the toll $p$  either too low or too high can result in network instability. Second, in the case study, a toll of approximately 5 \$/veh is identified as optimal for maximizing the lower bound of throughput.

\subsubsection{Impacts of variances of compliance rates}
\begin{figure}[htbp]
    \centering
    \begin{subfigure}{0.25\linewidth}
        \centering
        \includegraphics[width=\linewidth]{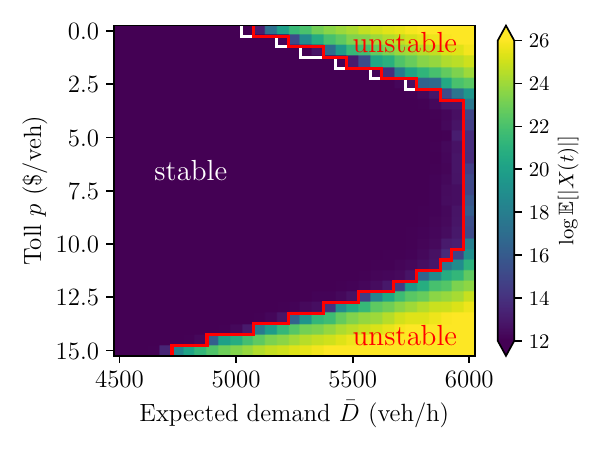}
    \caption{$\epsilon_{e_2}=0$.}
    \end{subfigure}
    \begin{subfigure}{0.25\linewidth}
        \centering
        \includegraphics[width=\linewidth]{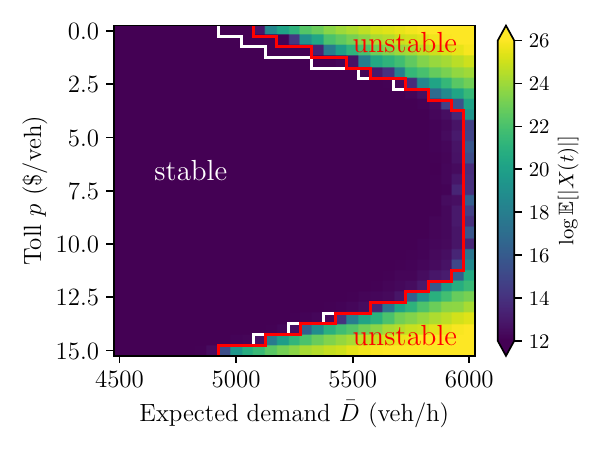}
    \caption{$\epsilon_{e_2}=0.2$.}
    \end{subfigure}
    \begin{subfigure}{0.25\linewidth}
        \centering
        \includegraphics[width=\linewidth]{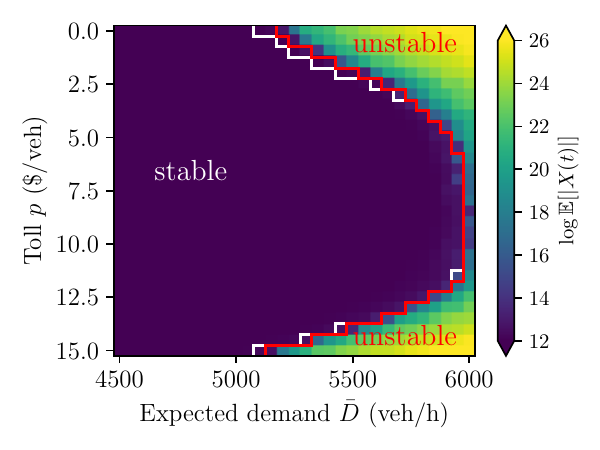}
    \caption{$\epsilon_{e_2}=0.4$.}
    \end{subfigure}
    \caption{Stability and instability regions under different $\epsilon_{e_2}$.}
    \label{fig_variance}
\end{figure}

Figure~\ref{fig_variance} illustrates the impacts of compliance rate variances by keeping the same $\bar{C}_{e_2}(x,p)$ but  selecting different $\epsilon_{e_2}$. From the upper right corners of Figures~\ref{fig_variance}(a)-(c), we can see the instability regions enlarge as $\epsilon_{e_2}$ increases. This  demonstrates that uncertainties in compliance rates may bring negative impacts on traffic management. From the lower right corners of Figures~\ref{fig_variance}(a)-(c), it is interesting to observe that, for the same level of demand, increasing tolls can stabilize a previously unstable network as $\epsilon_{e_2}$ increases. This result is reasonable since more uncertainties indicate higher tolls to persuade drivers to choose link $e_2$. 

More importantly, our white and red boundaries in Figures~\ref{fig_variance}(a)-(c) capture those necessary changes. This demonstrates that our developed theorems offer practical yet powerful tools for evaluating traffic systems without the need for extensive simulations.

\section{Future Work}
This work offers several potential avenues for extension. First, more sophisticated pricing strategies, such as stepwise tolls, could be explored. Second, it would be valuable to investigate conditions under which the stability criterion is both necessary and sufficient. Third, the approach could be expanded to accommodate more complex network structures.